\documentclass[prb,twocolumn,showpacs,10pt,nofootinbib]{revtex4}
\usepackage[dvips]{graphicx}
\usepackage{color,array,dcolumn}
\usepackage{amsmath}
\usepackage{amssymb}
\usepackage{slashed}
\numberwithin{equation}{section}

\usepackage{mathrsfs}
\allowdisplaybreaks{}

\begin{document}
\title{Charge carriers with fractional exclusion statistics in cuprates}
% \author{P.A. Marchetti,${}^1$ F. Ye,${}^2$ Z.B. Su,${}^3$  L. Yu${}^{4,3}$}
% \affiliation{${}^1$ Dipartimento di Fisica , INFN, I-35131 Padova, Italy\\
% ${}^2$ Institute for Advanced Study, Tsinghua University, 100084 Beijing,  China\\
% ${}^3$ Institute of Theoretical Physics, Chinese Academy of Sciences, 100190 Beijing, China\\
% ${}^4$ Institute of Physics, Chinese Academy of Sciences, 100190
% Beijing, China}
\author{P. A. Marchetti} \affiliation{Dipartimento di Fisica e
  Astronomia, INFN, I-35131 Padova, Italy}
\email{marchetti@pd.infn.it}
\author{F. Ye}
\affiliation{Department of Physics, Southern University of Science and
  Technology, Shenzhen 518055, China}
\email{yef@sustc.edu.cn}
\author{Z. B. Su}
\affiliation{Institute of Theoretical Physics, Chinese Academy of
  Sciences, 100190 Beijing, China}
\author{L. Yu} \affiliation{Institute
  of Physics, Chinese Academy of Sciences,and Beijing National
  Laboratory for Condensed Matter Physics, 100190 Beijing, China}
\affiliation{University of Chinese Academy of Sciences, 100049 Beijing, China}
\begin{abstract}
% We show that one can attribute consistently an exclusion statistics
%with parameter 1/2 to the charge carriers of the $t$-$J$ model in two dimensions, as it occurs in one dimension (1D). As in 1D, the no-double occupation constraint is
%   the origin of this fractional  statistics. With this
%  statistics many features of the hole-doped cuprates can be naturally
%  explained.
%Via coupling fermions to the Chern-Simons gauge field one can propose different schemes, in general with braid
%statistics, to implement the no-double occupancy constraint in the spin-charge decomposition for
%the two-dimensional (2D) $t$-$J$ model describing the physics of cuprates. All these choices should keep the Fermi statistics of physical holes, and, to be consistent with experiments, at large doping they should exhibit the standard Luttinger "large" Fermi volume.
We show
that in the $SU(2) \times U(1)$ spin-charge gauge approach we developed earlier one can attribute consistently an
exclusion statistics with parameter 1/2 to the spinless charge carriers of the $t$-$J$ model in two dimensions
(2D), as it occurs in
one dimension (1D). Like the 1D case, the no-double occupation constraint is at the origin of this
fractional exclusion statistics. With this statistics we recover a "large" Fermi volume  of holes
 at high dopings, close to that of the tight binding approximation. Furthermore, the composite nature
 of the hole, made of charge and spin carriers only weakly bounded, can provide a natural explanation
  of many unusual experimental features of the hole-doped cuprates.
\end{abstract}
\maketitle
\section{Introduction}
Despite continuous experimental advances it has not yet been
obtained an agreement on the interpretation of the low-energy
physics of  cuprates; see, $e.g.$, Ref. \onlinecite{uchida} for an
excellent, state-of-the-art review. A general consensus has been
achieved, however, that most of the relevant phenomena in hole-doped
materials can be derived from modeling the CuO planes in these
materials in terms of a two-dimensional (2D) $t$-$J$ model on the
copper sites with the Hamiltonian:
\begin{equation}
\label{eq:1}
H_{t-J}=\sum_{\langle i,j \rangle} P_G \left[ -t c^*_{i \alpha} c_{j \alpha} + J {\vec S}_i \cdot {\vec S}_j \right] P_G,
\end{equation}
where $\langle i,j\rangle$ denotes the nearest
neighbour (NN) sites,
%of the lattice,
$c_{i\alpha}$ the hole field operator with spin index
  $\alpha$ on site $i$, $P_G$ the Gutzwiller projection eliminating
double occupation,
and summation over repeated spin (and vector) indices is understood
hereafter. The Gutzwiller projected holes describe the
Zhang-Rice singlets \cite{zr} of cuprates.

A way of implementing the Gutzwiller projection is to apply to
fermions of the model a spin-charge decomposition formalism. It has
been pioneered by Anderson \cite{an} and Kivelson \cite{kiv} and it
is suggested for  cuprates by the rather different response of
charge and spin degrees of freedom in many experiments: One rewrites
the fermion field $c_\alpha$ as a product of a spinless holon field
$h$ carrying the charge degree of freedom and a spin 1/2 spinon
field $s_\alpha$ carrying the spin degree of freedom, imposing on
them a constraint reproducing the Gutzwiller projection. Due to this
decomposition an emergent (slave-particle) $U(1)$ gauge symmetry
appears, since holon and spinon fields can be multiplied by factors
with opposite phases leaving the original fermion field unchanged.
For this reason only slave-particle gauge-invariant fields are
physical and therefore neither the holon nor the spinon by
themselves are physical and they are strongly coupled by gauge
fluctuations. However, by gauge-fixing this gauge symmetry one can
consider gauge-dependent fields, which may be convenient in the
description of at least some momentum-energy range, as gluons and
quarks  in high-energy QCD, in spite of the fact that only mesons
and baryons are physical in strict sense. It makes sense then to
discuss  the statistics of holon and spinon fields and of the
"quasi-particle" excitations, in this generalized sense, that they
might generate in the low-energy limit. This is a key issue of this
article.
Notice that the spin-charge decomposition formalism does not prohibit a priori that the fermion field $c_\alpha$ describes an elementary excitation without a composite structure, as it happens e.g. for the meson field written in terms of quark fields in lattice QCD in the "super-confining phase", as discussed in Ref. \onlinecite{frobri}. Whether the fermion excitation described by $c_\alpha$ is composite or not is a dynamical question, it does not have a purely kinematical character as, on the contrary, the spin-charge decomposition does. Somewhat related considerations on the kinematical character of the spin-charge decompositions versus the dynamical character of their low-energy excitations can be found in Ref. \onlinecite{lee}.

 As rigorously shown in Ref. \onlinecite{fro}, spin-charge
decomposition can be achieved in the Lagrangian formalism by
coupling the original fermions with suitably chosen Chern-Simons
gauge fields. Furthermore, one may change the braid statistics of
holons and spinons, while still keeping the Fermi statistics of the
original holes. Different choices of Chern-Simons actions precisely
reflect these different braid statistics.

In one- and two-dimensions abelian braid statistics of particles
(or of field operators that create them) can be characterized by the
phase factor $e^{ \pm i(1-\alpha) \pi}$ acquired by their many-body
wave-function (or the product of equal-time field operators) when
one performs an oriented exchange among two of them, with $\alpha
\in [0,2)$ and $\pm$ referring to the two orientations (see, $e.g.$,
Ref. \onlinecite{wil}). Fermions(bosons) correspond to
$\alpha=0$(1), while excitations with $\alpha =1/2$ are called
semions.

All appropriate choices of holon and spinon statistics reproduce
exactly the correlation functions of the original fermion fields,
%model,
 as shown in
%some
the specific examples of slave bosons, slave fermions and slave semions
in Refs. \onlinecite{fro}, and using techniques developed there many
more schemes can be employed, $e.g.$, the slave anyons considered in
Ref. \onlinecite{ls} or a variant of the slave semion approach considered in
Ref. \onlinecite{msy98}, on which our subsequent discussion is
based. Slightly different roads follow, $e.g.$, the approach of Ref.
\onlinecite{wlmw}, as commented in this spirit in Ref.
\onlinecite{msy}.

Although all Chern-Simons choices are completely equivalent if
implemented exactly, as soon as one makes some mean-field-like
approximation they give rather different results\cite{msy}. It is
then crucial to understand which choice is better to perform
mean-field treatments. As discussed below, a key issue is the area
(2D volume) of the Fermi surface and that issue is in turn closely
linked to another form of statistics for the elementary excitations
of the model, $i.e.$, the $exclusion ~
statistics$.

The exclusion statistics was introduced by Haldane \cite{ha} to
generalize the Pauli exclusion principle. It can be characterized at
finite density
%(for $g \neq 1$ to have a Fermi surface)
by the average occupation of momenta at $T=0$ as follows \cite{wu}:
Consider a Fermi gas with fixed ``volume'' (in 2D later on called
area) enclosed by the Fermi surface and let $n_0$ be the
corresponding fermion density; we say that a particle obeys exclusion
statistics with parameter $g$ if the particle density with the same
Fermi ``volume'', denoted by $n_g$, satisfies
\begin{align}
\label{eq:2}
n_0=(1-g) n_g.
\end{align}
This implies that at a fixed momentum (neglecting other internal
degrees of freedom) a particle with exclusion statistics 1/2 can
have an occupation number twice that of a free fermion, so that the
``volume'' of its Fermi surface is half of that of a Fermi gas with
the same density. In a ``cartoon language'' one might say that a
semion in momentum space behaves like ``one half'' of a fermion.

For the statistics of ``quasi particle'' excitations of the 2D $t$-$J$ model and the
related Fermi surface area we have two sources of suggestions: The
solvable one-dimensional (1D) $t$-$J$ model and, under the assumption
initially made, the experiments on cuprates.

Concerning the 1D model the answer for the braid statistics is
unique: Both  holon and spinon fields and the related low-energy "quasi-particles" should be semions, $i.e.$, with braid
statistics parameter $1/2$, to reproduce in a suitable mean-field
treatment the correct scaling limit of correlation functions
obtained via Bethe ansatz or conformal field theory methods
\cite{haha,np}.
% Perhaps the reason underlying this statistics is that with this choice in a mean-field treatment  a spin flip appears when the holon hops and this permits to avoid a competition between  $t$ and $J$ terms, as sketched later on.

Furthermore, the holon in 1D has exclusion
statistics parameter $g=1/2$.
As shown in Ref.\onlinecite{ymsy}, in general there is no relation
between the braid and exclusion statistics in 1D. In fact one can introduce the braid statistics coupling
1D spinless fermions  to a Chern-Simons field and then perform its dimensional reduction. The
physical gauge-invariant field is obtained by adding to the fermion
field a gauge string and it obeys a braid statistics consistent with
the Chern-Simons coefficient. The Fermi points are shifted by the
gauge string but the Fermi 1D ``volume'' remains constant, hence the
corresponding low-energy excitations still obey an exclusion Fermi statistics.
A non-trivial exclusion statistics emerges if the fermionic fields
have a Luttinger interaction, a connection previously clearly stated
in Ref. \onlinecite{wy}. In the approach of Ref. \onlinecite{np} to
the 1D $t$-$J$ model, neglecting at first the coupling with spinons,
the holon at large scale behaves as a free U(1)
%spinless
semion;  spinons are described by Gutzwiller projected fermions in a
squeezed chain obtained by omitting the holon sites and at large scale
they form a semion gas with exclusion parameter $g=1/2$
described effectively by a Luttinger liquid theory. We then perform a field redefinition, eliminating the Fermi
surface for the spinon fields by suitably stripping away their gauge
strings and adding them to the holon in the correlation
functions of the physical hole. As result the 1/2 exclusion statistics
of spinons is transferred to holons;  holons then have
$g =\alpha=1/2$,
%Although in general in 1D there is no relation between braid and exclusion statistics, as shown in \cite{ym}, in the 1D $t$-$J$ model it is proved in \cite{np} that the Gutzwiller projection enforces the relation $g =\alpha$ for the holon,
so that the Fermi momentum of the U(1)
%spinless
semionic holon equals the Fermi momentum of the original spin 1/2
fermion treated in the tight binding approximation, in agreement
with the exact solution of the model. In fact in such exact solution
the Fermi points of the hole are in the position expected for a spin
1/2 fermion with the standard Pauli principle, consistently with the
Luttinger theorem, as extended to 1D Luttinger liquids in Ref.
\onlinecite{yoa}. Since in the spin-charge decomposition spinons
with the above redefinition don't have a Fermi surface while holons
are spinless, that result is correctly recovered by the 1/2
exclusion statistics of holons.

We now turn to the suggestion
%for 2D
coming from experiments on cuprates.  In overdoped materials the
Fermi surface seen in ARPES is close to that obtained in a
tight-binding approximation of a $t$-$t'$-$J$ model and satisfies
the standard Luttinger theorem. (The introduction of a next nearest
neighbor  (NNN) hopping parameter $t'$, and possibly a NNNN $t''$,
in the formalism discussed here is straightforward and it does not
change the qualitative features, so it will not be elaborated
anymore). To reproduce this result in the spin-charge decomposition
formalism we have two natural options: Either the spin 1/2 spinon is
fermionic with Fermi surface and the holon is a hard-core boson, as
in the slave-boson approach (see, $e.g.$. Ref. \onlinecite{lee}), or
the spinless holon has Fermi surface with exclusion statistics
parameter 1/2, hence with the same Fermi surface of spinons of the
slave-boson approach, while the spinon has no Fermi surface.

We see that, if both spinon and holon are semions, the second case would
be a close analogue to what happens in 1D.
%It turns out that as in 1D in mean-field a spin flip arises in the links where the holon hops, thus avoiding the competition between $t$ and $J$ terms.
 We remarked above that in 1D
the natural condition for the appearance of exclusion statistics is the
Luttinger interaction; analogously on general grounds we proved in
Ref. \onlinecite{ym} that in 2D we have $g = \alpha$ if the
original fermionic system without Chern-Simons
coupling has Hall conductivity 1/2$\pi$ and is incompressible. The main
goal of this paper is to show that indeed these conditions can be
satisfied for the holon in the 2D $t-J$ model and the second approach
considered above to get the correct Fermi area can be consistently
implemented, sketching also a derivation from it of some consequences
for cuprates.

\section{The $SU(2) \times U(1)$ spin-charge gauge approach}
To implement a semionic decomposition of the hole in the 2D $t$-$J$
model we start by making use of the following theorem \cite{fro,np}.

{\it Theorem}: We embed the lattice of the 2D $t$-$J$ model in a
3-dimensional space, denoting by $x=(x^0,x^1,x^2)$ coordinates of
the corresponding 2+1 space-time, $x^0$ being the euclidean time. We
couple  fermions of the $t$-$J$ model to a $U(1)$ gauge field,
$B^\mu$, gauging the global charge symmetry, and to an $SU(2)$ gauge
field, $V^\mu$, gauging the global spin symmetry of the model, and
we assume that the dynamics of the gauge fields is described by the
Chern-Simons actions $-2 S_{c.s.}^{U(1)} (B) + S_{c.s.}^{SU(2)} (V)$
with:
\begin{align}
\label{eq:3}
   S_{c.s.}^{U(1)} (B) =&  \frac{1} {4 \pi i} \int d^3 x \epsilon_{\mu\nu\rho}
   B^\mu \partial^\nu B^\rho (x), \nonumber\\
   S_{c.s.}^{SU(2)} (V) =& \frac{1} {4\pi i} \int d^3 x {\rm Tr} \epsilon_{\mu\nu\rho}
   [V^\mu \partial^\nu V^\rho + \frac{2}{3} V^\mu V^\nu V^\rho](x), \nonumber\\
\end{align}
% \begin{displaymath}
% \label{eq:3}
%    S_{c.s.}^{U(1)} (B) =  \frac{1} {4 \pi i} \int d^3 x \epsilon_{\mu\nu\rho}
%    B^\mu \partial^\nu B^\rho (x),
% \end{displaymath}
% \begin{equation}
%    S_{c.s.}^{SU(2)} (V) = \frac{1} {4\pi i} \int d^3 x {\rm Tr} \epsilon_{\mu\nu\rho}
%    [V^\mu \partial^\nu V^\rho + \frac{2}{3} V^\mu V^\nu V^\rho](x),
% \label{eq:4}
% \end{equation}
where $\epsilon_{\mu\nu\rho}$ is the Levi-Civita anti-symmetric
tensor in three dimensions.  Then the spin-charge (or $SU(2) \times
U(1)$) gauged model so obtained is exactly equivalent to the
original $t$-$J$ model.  In particular the spin and charge invariant
correlation functions of the fermion fields $c_{j \alpha}$ of the
$t$-$J$ model are exactly equal to the correlation functions of the
fields $ \exp(-i \int_{\gamma_j} B) P[\exp(i \int_{\gamma_j}
V)]_{\alpha \beta} c_{j \beta}$, where $c$ denotes now the fermion
field of the gauged model, ${\gamma_j}$ a string at constant
euclidean time connecting the point $j$ to infinity and $P(\cdot)$
the path-ordering, which amounts to the usual time ordering
$T(\cdot)$, when ``time'' is used to parametrize the curve along
which one integrates. One can view the result of this theorem as an analogue of the construction
 of composite fermions in Jain's approach to the Quantum Hall Effect \cite{jain}. In that case
 magnetic vortices with even quantum flux (depending on the filling) are bound to the electron
 and the resulting composite entity is still a fermion, dubbed composite fermion; in the present case
  the electron of the $t$-$J$ model is bound to a charge-vortex of flux -1/2 and a spin-vortex of flux 1/2,
   while the resulting entity still being a fermion.

Notice that, contrary to what one might naively think, although the
Chern-Simons actions individually explicitly break the parity and time-reversal
symmetries, the particular combination considered above
still preserves explicitly these two symmetries.

We now rewrite the hole field $c$ of the gauged model as a product
of a charge 1 spinless fermion field $h$ and a neutral spin 1/2
boson field $\tilde s_\alpha$: $c_\alpha=h^* \tilde s_\alpha$. Then
we identify $ \exp[i \int_{\gamma_j} B]h_j$ as the holon and $
P(\exp[i \int_{\gamma_j} V])_{\alpha \beta} \tilde s_{j \beta}$ as
the spinon fields. The Chern-Simons coupling automatically ensures
that both corresponding field operators obey semionic braid
statistics.  The holon $h$ being spinless implements exactly the
Gutzwiller constraint due to the Pauli principle. Furthermore, if
the constraint $ (P(\exp[i \int_{\gamma_j} V])_{\alpha \beta} \tilde
s_{j
  \beta})^\dagger (P(\exp[i \int_{\gamma_j} V])_{\alpha \beta'} \tilde
s_{j \beta'})=\tilde s^*_\alpha \tilde s_\alpha=1$ is imposed, as $
c^*_\alpha c_\alpha=1-h^* h$ we see that $(\exp[i \int_{\gamma_j}
B]h_j)^*(\exp[i \int_{\gamma_j} B]h_j)=h_j^* h_j$ is just the
density of empty sites in the model, corresponding to the Zhang-Rice
singlets.

The "charge-flux" associated to the electrons produces a $\pi$-flux
phase for every plaquette, plus vortices centered on the empty
sites, $i.e.$ on the holon positions. More precisely, introducing a
Coulomb gauge-fixing $\partial_\mu B^\mu=0$ for the $U(1)$ charge
gauge symmetry, one finds

%Let us come back to the action of the 2D gauged $t$-$J$ model and
%integrate out $B^0$; we obtain the constraint (with
%$\mu,\nu=1,2, z \in \mathbf{R}^3$):
%\begin{eqnarray}
%\label{eq:5}
%\epsilon_{\mu\nu}\partial^\mu B^\nu(z)=\pi [\sum_j (1-h^*_jh_j)(z^0)\delta^{(2)}(\vec z-j)].
%\end{eqnarray}
%Imposing the Coulomb gauge-fixing $\partial_\mu B^\mu=0$ one gets
$B^\mu(z)=\bar B^\mu + \bar b^\mu(z)$, where
% $\bar B^\mu$ introduces a
%$\pi$-flux phase, $i.e.$, $\exp(i\int_{\partial p} \bar B)=-1$ for
%every plaquette $p$ and
\begin{eqnarray}
\label{eq:6}
\bar b^\mu(z)=\frac{1}{2}[\sum_j \partial^\mu \arg(\vec z-j)( h^*_jh_j)(z^0)].
\end{eqnarray}
and we can choose
%We  recognize $ \partial^\mu \arg(\vec z-j)$ as the vector potential
%of a vortex centered at the holon position $j$, $i.e.$, centered at
%an empty site of the $t$-$J$ model. Vortices in Eq. (\ref{eq:6})
%appear in the charge $U(1)$ group and are  responsible for the
%semionic nature of  holons.  We at first neglect these
%charge-vortices and choose for $\bar B$ the staggered $\pi$ flux
%form:
\begin{eqnarray}
\label{eq:7}
\bar B_{\langle i,i \pm \vec e_x \rangle} = \pm \pi/4, \bar B_{\langle i,i \pm \vec e_y \rangle} = \mp \pi/4,
\end{eqnarray}
where $i$ is a site of the even N\'{e}el sublattice and
  $\vec e_x, \vec e_y$ the two unit vectors along the link
  directions.
  We  recognize $ \partial^\mu \arg(\vec z-j)$ as the vector potential
of a vortex centered at the holon position $j$, $i.e.$, centered at
an empty site of the $t$-$J$ model. Vortices in Eq. (\ref{eq:6})
appear in the charge $U(1)$ group and are  responsible for the
semionic nature of  holons.

Neglecting at first these charge-vortices, through Hofstadter mechanism the $\pi$-flux converts the spinless
holon field $h$ into a pair of ``Dirac fields'' in the magnetic
Brillouin zone (BZ), with pseudospin indices corresponding to the
two N\'{e}el sublattices and two ``small FS'' centred at $(\pm
\frac{\pi}{2},\pm \frac{\pi}{2})$.
% given the form of $\bar B$ as in Eq.~\eqref{eq:7}.
 If we reinsert the
charge-vortices, and assume for the corresponding semionic holons an
exclusion statistics 1/2, then these holons have the same FS of the
fermionic spinons of the slave-boson approach in the $\pi$-flux
phase \cite{lee}, and the same dispersion: $\omega_h \sim 2t
[\sqrt{\cos^2 k_x + \cos^2 k_y }- \delta]$, where $\delta$ denotes
the density of empty sites, corresponding in  cuprates to the
in-plane doping concentration. When these holons are coupled to
spinons through the self-generated slave-particle gauge field, as a
consequence of the Dirac structure of the holons the  resulting
holon-spinon bound state generated in the low-energy limit  exhibits
Fermi arcs qualitatively consistent with those found in ARPES
experiments in the pseudogap ``phase'' in cuprates \cite{msy04}. The
underlying FS for the hole is  modified $w.r.t.$ the holon FS by the
spinon gap proportional to $\delta^{1/2}$ discussed later.
Furthermore, using the techniques of
Refs.\onlinecite{msy04}, \onlinecite{mg} and assuming  a doping
independent renormalization of the spinon gap, the fraction of BZ
enclosed can be for all dopings approximately $\delta/2$, where the
factor 1/2 comes precisely from the 1/2 exclusion statistics
\cite{mysy}.
% Furthermore, with the inclusion of a $t'$ term the location of the Fermi arcs turn out to be essentially on the FS of the tight-binding of the $t$-$t'$ model at the same doping \cite{mysy}. }

%Although it is not the main concern of this paper
%Let us now briefly sketch the parallel analysis for  spinons, as it
%allows to better understand the importance of the exclusion
%statistics for holons. Furthermore, some details will be needed in
%the proof of exclusion statistics which will be relevant for the
%mechanism of superconductivity outlined below. As discussed in Ref.
%\onlinecite{m}, we proceed in analogy with the 1D treatment by the
%following steps:
We can now use the $SU(2)$ gauge freedom to rotate the spinons $\tilde s$
to a configuration $\tilde s^m$, depending on the holon
configuration, optimizing ``on average'' the holon-partition
function in that spinon background, in a Born-Oppenheimer approximation. In this configuration
 spinons are antiferromagnetically ordered along the magnetization direction of the undoped model,
 which we arbitrarily fix along $z$. There is in addition a spin flip on the sites where holons are
 present, also for the final
site of a hopping link of holons, at the time of hopping.  Above a crossover temperature $T^*$ we find that
$\tilde s^m$ involves also a phase factor cancelling the contribution of
$\bar B$ in the loops of hopping links of holons, so that the hopping
holons feel an approximately zero flux \cite{msy05}.  Assuming the
exclusion statistics with parameter 1/2 for the holon (to be proved in
the next Section), the disappearance of the $\pi$-flux implies that the Hofstadter mechanism does not
 hold anymore and above
$T^*$ one recovers  for  holons the ``large FS'' of the
tight-binding approximation. In particular the fraction of BZ
enclosed is of order $(1 + \delta)/2$ and the factor 1/2 comes
precisely again from the 1/2 exclusion statistics. This FS will be
inherited by the physical hole as a holon-spinon bound
state\cite{msy05}, and, with the addition of a $t'$ term and a
renormalization of the spinon gap, it is in approximate agreement
with the FS observed in ARPES in the "strange-metal" region of the
phase diagram of the cuprates \cite{mysy}. For this reason we call
pseudogap (PG) for the $t$-$J$ model in our approach the region
below $T^*$ and strange metal (SM) the region above it. Actually,
although only qualitatively, the above results on the FS have been
proved in Refs. \onlinecite{msy04}, \onlinecite{msy05} in a rough
approximation in which the 1/2 Haldane statistics for the holon was
assumed, but, somewhat inconsistently, the semionic nature of the
holon field was not taken into account, keeping, on the other hand,
the treatment of the spinon consistent with this approximation.
%We expect that a more careful account of the semion nature of the holon don't change the above features of the FS, and it is presently under investigation.
As discussed in the introduction the proof of the Haldane statistics of the holon is the main aim
 of this paper, but this proof needs some more details on the spin-charge approach that we now provide.
 At the end of the paper some results obtained with this approach, including a non-BCS mechanism for
 superconductivity, are outlined, making also contact with experiments in cuprates.

%1) We gauge-fix the $SU(2)$ gauge symmetry by imposing (even in the
%presence of holons) the requirement that  spinons $\tilde s$ are
%antiferromagnetically ordered, by setting
%\begin{eqnarray}
%\label{eq:8}
%\tilde{s}_j = \sigma_x^{|j|} \begin{pmatrix}
% 1\\0
%\end{pmatrix},
%\end{eqnarray}
%where $|\vec{j}|=j^1+j^2$ denoting the parity of the
%  lattice $\vec{j}$. Let us remark that Eq. (\ref{eq:8}) being  a
%gauge-fixing it does not imply long-range antiferromagnetism for the
%physical spins, as it will be clear in the following discussion.
Having used the $SU(2)$ gauge freedom to rotate spinons to the
"optimal" configuration $\tilde s^m$, we need to integrate the
$SU(2)$ gauge field $V$ over all its configurations. Therefore, we
split the integration over $V$ into an integration over a field
$\bar V$, satisfying the Coulomb gauge-fixing $\partial_\nu \bar
V^\nu=0$ with $\nu=1,2$ and its gauge transformations expressed in
terms of an $SU(2)$-valued scalar field $U$, $i.e.$, $V^\mu=
U^\dagger \bar V^\mu U+U^\dagger\partial^\mu U$. Notice that $
P[\exp(i \int_x^y V)]=U^\dagger_y P[\exp(i \int_x^y \bar{V})]U_x$.

%3) We consider a Born-Oppenheimer approximation
%for the spinons in the presence of the holons. We find a configuration of $g$, depending on the holon
%configuration, optimizing ``on average'' the holon-partition
%function in that $g$ background, and around that configuration,
%$g^m$, we consider spinon fluctuations, $i.e.$, we write $g=U g^m$,
%with $U \in SU(2)$ describing fluctuations around the optimal
$U$ describes fluctuations of  spinons around the "optimal"
configuration and can be written as:
\begin{eqnarray}
\label{eq:9}
U= \begin{pmatrix}
 s_1 &-s_2^*\\ s_2 & s_1^*
\end{pmatrix},
\end{eqnarray}
with $s$ satisfying the constraint:
\begin{eqnarray}
\label{eq:10}
s^*_{\alpha j} s_{\alpha j} =1.
\end{eqnarray}
We will call in the following $s_\alpha$ again ``spinons''. Up to now no approximation has been
made and the model in terms of $h$, $s$ and $\bar V$ is still equivalent to the original $t$-$J$ model.
 However, due to the optimization procedure on the spinon, we expect that the configurations of $U$ are
 dominated by small fluctuations around identity.
As discussed in
Ref. \onlinecite{msy}, the spin flip due to the $SU(2)$ gauge freedom in
  $\tilde s^m$ allows a simultaneous optimization in terms of the spinon $s$ both of the $t$ and the $J$
   term in the Born-Oppenheimer approximation
considered above, because (neglecting $\bar V$) $s$ appears in the form $s^*_{\alpha i} s_{\alpha j}$
on links in the $t$ term and as $|\epsilon_{\alpha \beta} s_{\alpha i} s_{\beta j}|^2$ and the identity
 $|s^*_{\alpha i} s_{\alpha j}|^2 + |\epsilon_{\alpha \beta} s_{\alpha i} s_{\beta j}|^2 =1$ holds.
  This phenomenon occurs also in 1D and might be the origin of the good mean field approximation for the semionic statistics.

We now briefly discuss a ``mean field'' approximation essentially based on the conjecture that the fluctuations
 $U$ are small.
If we neglect fluctuations $U$ in the calculation of $\bar V$ (up to
an irrelevant field-independent term) one gets ( with $\mu=1,2$):
\begin{eqnarray}
\label{eq:11}
\bar V^\mu(z)= -\frac{1}{2} \sum_j (-1)^{|j|} \partial^\mu\arg(\vec z-j) h^*_jh_j(z^0) \sigma_z.
\end{eqnarray}

We recognize in the term $(-1)^{|j|} \partial^\mu\arg(\vec z-j)$ of
Eq.(\ref{eq:11}) the vector potential of a vortex centered at the
holon position $j$, with vorticity (or chirality)
  depending on the parity of $|j|=j_x+j_y$.  We call these vortices
antiferromagnetic (AF) spin vortices, since they record in their
vorticity the N\'{e}el structure of the lattice. Hence they are
still a peculiar manifestation of the AF interaction, like the more
standard AF spin waves. As one can see,
  they are the topological excitations of the  $U(1)$
  subgroup of the original $SU(2)$ spin group unbroken in the AF
  phase, along the spin direction $z$ of the magnetization.

These vortices are of purely quantum origin, since, like in the
Aharonov-Bohm effect, they induce a topological effect far away from
the position of the holon itself, where their classically visible
field strength is supported. Hence in this approach the empty sites
of the 2D $t$-$J$ model, mimicking the Zhang-Rice singlets and
corresponding to the holon positions, are  cores of the AF spin
vortices, quantum distortions of the AF spin background. These
vortices have no analogue in the slave-boson approach and in our
approach are responsible for both short-range AF order, which we now outline
 since it will be used in the proof of Haldane statistics, and a new
pairing mechanism leading to superconductivity, sketched in the final section,
referring to Ref. \onlinecite{mfsy} for details.
%Here we present the main ideas, deferring more technical formulas to Appendix A.
Semionic holons dressed by AF
spin vortices are similar to semionic holons in 1D with attached
 a spinon-derived "spin string"; the role of kinks as topological defects
in 1D is replaced by vortices in 2D.

We can write the total action of the system as a sum of a "spinon action" $S_s$ and a "holon action" $S_h$.
For our purpose of the spinon action it is enough to know \cite{msy98} that  in the long wavelength
continuum limit it is given by a $O(3)$ non-linear $\sigma$ model (in $CP^1$ form) for spinons describing
the continuum limit of the undoped Heisenberg model %which can be put in $CP1$ form with a slave-particle gauge field $A_\mu = s_a \partial_\mu s_\alpha$,
 with an additional coupling between spinons $s$ and the AF spin vortices :
\begin{eqnarray}
\label{eq:14}
\int d^3x (\bar V^\mu \bar V_\mu)(x) s^*_\alpha s_\alpha (x).
\end{eqnarray}
A quenched average, $\langle \cdot \rangle$, over positions of
centers for spin-vortices yields the following estimate
\cite{msy98}: $\langle \bar V^\mu \bar V_\mu\rangle \approx \delta
|\log \delta|$. Hence the term (\ref{eq:14}) provides a mass-gap to
spinons,  converting the long-range AF order of the Heisenberg
model, corresponding to zero doping, to short-range AF order at
finite dopings; therefore, spinons $s$ have no FS. The spinon system behaves as a spin liquid
since spinon confinement is avoided by  the interaction with the gapless holons. However, in
spite of the presence at lattice level of a Chern-Simons term which turns spinons into semions,
 in the mean-field long-wavelength limit considered involving the coupling to the holons via AF
  spin vortices,  it is not a chiral spin liquid and  spinons $s$ in the low-energy limit can
   be considered as spin 1/2 hard-core boson quasi-particles excitations. Although not confined,
    in the entire system spinons are weakly bound to holons and anti-spinons by slave-particle gauge
    fluctuations to form the physical composite holes and magnons, respectively.

By making  in Eq. (\ref{eq:14}) a mean-field approximation for $ s^*_\alpha s_\alpha (x)$,
instead of what was previously considered for $\bar V^\mu \bar V_\mu(x)$, we obtain
the term
 \begin{eqnarray}
 \label{eq:15}
   \langle s^*_\alpha s_\alpha \rangle \sum_{i,j} (-1)^{|i|+|j|}\Delta^{-1} (i - j) h^*_ih_i h^*_jh_j ,
\end{eqnarray}
where $\Delta$ is the 2D Laplacian. In the static approximation for
holons Eq. (\ref{eq:15}) describes a 2D lattice Coulomb gas with
charges $\pm 1$ depending on the N\'{e}el sublattices. In particular
the interaction is attractive between holons in opposite N\'{e}el
sublattices, with maximal strength for nearest neighbor sites, along
the lattice directions with a $d$-wave symmetry. Putting back
coefficients one finds that the coupling constant of this
interaction is $J_{eff}=J(1-2\delta)\langle s^*_\alpha s_\alpha
\rangle$, which decreases with increasing doping.  For 2D Coulomb
gases with the above parameters, pairing appears below a temperature
$T_{ph} \sim J_{eff}$. Hence the charge-pairing originates from the
attraction between AF spin vortices with opposite chirality, eventually leading to
 superconductivity as sketched in the final section.

We write now explicitly the holon action, $S_h$, since its expression is needed in the proof of exclusion statistics.  In PG region $S_h$ can be written as:
\begin{widetext}
\begin{align}
\label{eq:12}
S_h=&\int dx^0 \sum_j [h^*_j [\partial_0 -i b_0(j)-2 t \delta] h_j ]- h^*_j h_j( \sigma_x^{|j|}[U^\dagger_j \partial_0 U_j +i v_0 \sigma_z] \sigma_x^{|j|})_{11} \nonumber \\&- \sum_{\langle i,j \rangle} t h^*_j
 \exp
  \left(
  i \bar B_{\langle i,j \rangle}-i\int_{i}^{j} b\right)h_i
  \left[
  \sigma_x^{|i|} U^\dagger_i \exp\left(i \int_{i}^{j} v \sigma_z\right) U_j\sigma_x^{|i|}\right]_{11}- 2S_{c.s.}^{U(1)}(b) + 2 S_{c.s.}^{U(1)} (v).
\end{align}
\end{widetext}
The fact that $\sigma_x$ has the same power at both ends of a
hopping link is due to the spin-flip generated by $\tilde s^m$. $b$ is a gauge field of the $U(1)$-charge
group and $v$ is a gauge field of the $U(1)$ subgroup of the spin group $SU(2)$  previously selected
by choosing the directions of $\tilde s^m$. The factor 2 in the Chern-Simons action is due to a normalization
 needed passing from $SU(2)$ to
its $U(1)$ subgroup. Integrating over
$b_0$ and $v_0$ one reproduces the previous description in terms of
 $\bar b_\mu$ and $\bar V_\mu$. Notice that since coefficients of the Chern-Simons
terms for $b$ and $v$ have opposite sign,  at this stage the parity (P) and time-reversal (T)
symmetries are still explicitly preserved.   Formally this can be seen by rewriting the gauge
fields in the combinations $b {\bf 1} +v \sigma_3$ and $b {\bf 1} -v \sigma_3$, where ${\bf 1}$
is the $2 \times 2$ identity matrix.  Holons are coupled only to $b {\bf 1} -v \sigma_3$,
so integrating  $b {\bf 1} +v \sigma_3$  from the Chern-Simons one gets  a delta function
 for the field strength of $b {\bf 1} -v \sigma_3$ which is P and T invariant.
  We argue, however, that the continuum limit should not be taken considering simultaneously the
  coupling of holons to $b$ and $v$, but firstly only to $b$, to make the holon field $U(1)$charge-gauge
   invariant and to enforce the semionic statistics, and only afterwards introducing the coupling
   with the spin degrees of freedom.

 To summarize in words, in PG $S_h$ describes fermionic lattice holons in the presence of  $\pi$ flux
  per plaquette with attached  charge-vortex generated by $b$ that turn them into semions,
  interacting with spinons $s$ and the AF spin vortices described by  $\bar V$. In SM the $\pi$
  flux in the hopping is suppressed.

\section{The 1/2 exclusion statistics of holons}
Having explained the relevance for self-consistency of the exclusion statistics 1/2 for
the holon in the spin-charge gauge approach to  cuprates, in this
Section we turn to its proof.

\subsection{The braid-exclusion statistics relation}
A key ingredient of the proof is the result contained in
Ref. \onlinecite{ym}, connecting braid and exclusion statistics under
some conditions, that we now sketch.

Consider a planar Hall system consisting of fermions in a
thermodynamically large domain with a boundary; it is well known that
there are chiral edge modes on the sample boundary leading to a boundary
current. We then couple the system to a Chern-Simons field $b_\mu$, defined in the whole space-time, with
coupling strength $\alpha$, while keeping fixed the chemical potential
$\mu$. We denote by $N(\mu,\alpha)$ the number of particles contained in
the considered domain with Chern-Simons coupling $\alpha$. The
Lagrangian (in real time) reads
\begin{equation}
\label{eq:16}
\mathcal{L}=\mathcal{L}_M - b_{\mu}J^{\mu} + \frac{1}{4\pi\alpha}\int
  d^3x \epsilon^{\mu\nu\lambda}b_{\mu}\partial_{\nu}b_{\lambda},
\end{equation}
where $J^{\mu}$ is the current density.  The exact form of the
Lagrangian $\mathcal{L}_M$ of  fermions is not so important, and it is
only required to provide a non-vanishing Hall conductance
$\sigma_H$.

By differentiating Eq.~(\ref{eq:16}) $w.r.t.$ $b_\nu$($\nu=1,2$),
one obtains the following relation between the current and the
``electric'' field: $\vec{E}\equiv \partial_0 \vec b-\vec\nabla b_0=
2\pi\alpha (\hat{z}\times \vec{J})$, where $\hat{z}$ is the unit
vector perpendicular to the plane of the system.  Thus a current
$I(\mu)$ flowing on the boundary leads to an electric field normal
to the boundary.

In the scaling limit the fermion Hall system
contributes a Chern-Simons term to the gauge effective action with
coefficient $-\sigma_H/2$ in the bulk (plus a term localized at the boundary by gauge-invariance).
 Taking into account this contribution in
random phase approximation (RPA) leads to an effective Chern-Simons
coupling for the $b$ field: $\tilde{\alpha}\equiv
\alpha/(1-2\pi\sigma_{H}\alpha)$. The above electric field generated
by the boundary current implies a jump $\Delta V=-2\pi\tilde{\alpha}
I(\mu)$ of the scalar potential across the sample boundary. Then the
change of the free energy $\mathcal{F}$ due to the Chern-Simons
coupling reads
\begin{equation}
\label{eq:17}
\mathcal{F}(\mu,\alpha) = \mathcal{F}(\mu,0) - 2\pi\tilde{\alpha}I(\mu)
  N(\mu,0).
\end{equation}
Differentiating both sides of Eq. (\ref{eq:17}) with respect to
$\mu$, we obtain
\begin{align}
\label{eq:18}
N(\mu,\alpha) =& N(\mu,0) +2\pi\tilde{\alpha} \frac{\partial I}{\partial
  \mu} N(\mu,0) \nonumber\\
&+ 2\pi\tilde{\alpha}I(\mu)\frac{\partial N(\mu,0)}{\partial\mu},
\end{align}
where $\partial I/\partial \mu=\sigma_H$ and $\partial N/\partial \mu$
is proportional to the compressibility. For an incompressible liquid, one obtains
\begin{equation}
\label{eq:19}
N(\mu,0)=(1-2\pi\sigma_H\alpha)N(\mu,\alpha).
\end{equation}
Since  $\mu$ was kept invariant, by comparing Eq.~(\ref{eq:19}) with
Eq.~(\ref{eq:2}) one concludes that anyons of the system described
in Eq.~(\ref{eq:16}) obey an exclusion statistics with parameter $g=
2 \pi \sigma_H \alpha$. In particular, if $ 2 \pi \sigma_H =1$ we
have $g=\alpha$.

Let us now come back to our holon system. The above argument shows
that if holons without Chern-Simons coupling to $b$ have a Hall
conductivity $1/ 2 \pi$ and the system is incompressible, then the
semionic holons obtained by coupling with $b$ obey exclusion statistics 1/2, as we would like to prove.

\subsection{The free holons}
In the ``holon action'' Eq. (\ref{eq:12}) if no further
approximations are made the holon density remains $\delta$, since
the Gutzwiller projection is still exactly implemented by $U \in
SU(2)$ with the constraint Eq.~(\ref{eq:10}) being satisfied.  In
particular when $\delta =0$ the holon density vanishes, correctly
reproducing the vanishing density of Zhang-Rice singlets at
half-filling.

However, in the large-scale continuum limit we have seen that,
thanks to the interaction with  AF vortices, the spinon $s(x)$ is
gapped. It implies that in this limit the constraint
Eq.~(\ref{eq:10}) is not fully satisfied, as the spinon mass gap is
incompatible with it, so the Gutwiller projection is not anymore
exactly implemented. To understand the situation let us first
consider the free holons without coupling to  spinons and the
Chern-Simons fields $b$ and $v$, while still keeping fixed the
chemical potential. The corresponding action is given by
 \begin{widetext}
\begin{eqnarray}
\label{eq:12a}
&S^0_h=\int dx^0 \sum_j h^*_j (\partial_0 - 2 t \delta) h_j - \sum_{\langle i,j \rangle} t h^*_j
 \exp[i \bar B_{\langle i,j \rangle}]h_i
\end{eqnarray}
\end{widetext}

Due to the staggered $\pi$-flux implemented by $ \bar B_{\langle i,j
\rangle}$, we divide the square lattice into two sublattices, A(even
sites) and B (odd sites). On these sublattices, the annihilation
operators of holons are denoted by $ h^a$ and $ h^b$, respectively.
Let's choose a unit cell with  A and B sites along the
$x$-direction, then the Hamiltonian corresponding to the free holon
action Eq.~(\ref{eq:12a}) can be recast in a quadratic form, with a
matrix in the momentum space given by:
\begin{align}
\label{eq:20a}
H(\vec k)= 2t\begin{pmatrix}
 0 & \cos k_x + i \cos k_y\\ \cos k_x-i \cos k_y& 0
\end{pmatrix}.
\end{align}
In Eq. ~(\ref{eq:20a}), the momentum $k$ only takes values in the
range $[-\pi,\pi] \times [-\pi/2,\pi/2]$, which is a half of the
original BZ. One can easily see that it describes two massless Dirac
double-cones with vertices at $(\pm \pi/2, \pi/2)$. Shifting the two
Dirac nodes to the origin in $k$-space, inserting the chemical
potential $\mu= 2 t\delta$ and taking the continuum limit, we see
that the corresponding continuum fields are described by massless
Dirac fields with two flavors, corresponding to the two
double-cones. The two upper bands of the double-cones are filled up
to energy $ 2 t \delta$, hence even at $\delta =0$ the lower bands
of the two Dirac double-cones are filled, so that the holon density
no more vanishes even in the half-filling case.

The lower bands are thus an artifact produced by the violation of
the constraint Eq.~(\ref{eq:10}) introduced when we treat in mean
field Eq.~(\ref{eq:14}).

%Keeping fixed the chemical potential was required in the introduction of the Chern-Simons coupling discussed in ~(\ref{eq:16}).

Since spinons are gapped, the Gutzwiller constraint is relaxed in
the large-scale continuum limit. Going back to the lattice model
with no-double-occupation constraint ignored temporarily, one
expects that at half filling with $\delta = 0$ the number of holons
equals the number of {\it unprojected} holes, hence one expects the
holon number is 1 per site on average. If these holons were fermions
obeying Fermi statistics, both upper and lower bands would be
completely filled, which is at odds with the previous half-filling
result obtained from the free holon Lagrangian.

This would lead to an inconsistency in the above continuum limit.
However, if  holons satisfy the semionic exclusion statistics with
$g = 1/2$, at half filling they fill the lower bands leaving the
upper bands empty to give a density 1 on average. Since these
semionic holons in the lower bands are a result of relaxing the
Gutzwiller projection, they are ``spurious'' and describe the singly
occupied sites in the original unprojected lattice model. When the
doping holes are introduced in the $t$-$J$ model, the corresponding
``physical'' holons partially fill the upper bands and are
responsible for the low energy physics. Although the spurious lower
band holons are not directly relevant to the low energy physics in
the scaling limit, they are responsible for the 1/2 exclusion
statistics when coupled to the $U(1)$ statistical field $b$, making
the theory self-consistent, as we prove below. Before closing this
subsection, we emphasize that one should be careful not introducing
an unphysical coupling of the ``spurious'' holons in the lower bands
with spinons, so that the density of ``physical''
holons coupled to spinons still correctly vanishes at $\delta = 0$.

\subsection{Hall conductivity of "spurious" holons}
According to the strategy outlined in subsection A we now compute
the Hall conductivity of the holon system without Chern-Simons
couplings. If we look at the corresponding ``holon action'' in
Eq.~(\ref{eq:12}), we see that for sites in the A sublattice and
links starting from the A sublattice the coupling with  spinons and
the $v$ field is of the form $(U^\dagger_j \partial_0 U_j+ iv_0
\sigma_z)_{11}$ and $[ U^\dagger_i \exp(i \int_{i}^{j} v \sigma_z)
U_j]_{11}$, whereas for the B sublattice the corresponding terms are
$(U^\dagger_j \partial_0 U_j + iv_0 \sigma_z)_{22}$ and $[
U^\dagger_i \exp(i \int_{i}^j v \sigma_z) U_j]_{22}$. Since
$(U^\dagger_j \partial_0 U_j + iv_0 \sigma_z)_{22}=(U^\dagger_j
\partial_0 U_j+ iv_0 \sigma_z)_{11}^*$ and
$[ U^\dagger_i \exp(i \int_{i}^{j} v \sigma_z) U_j]_{22}=[
U^\dagger_i \exp(i \int_{i}^{j} v \sigma_z) U_j]_{11}^*$, the action
is not invariant under time-reversal, but is invariant under
time-reversal combined with  interchange of the two N\'{e}el
sublattices realized by parity transformation with respect to a line
in the dual lattice. This can be intuitively understood since the time-reversal operation reverses
the chirality of the spin-vortices described by $v$, but  an exchange of the N\'{e}el
sublattices also does the same job.
%This suggests that this system has a ${\bf Z_2}$ topological order.

As well known \cite{re}, to compute the Hall conductivity of
massless Dirac fields we need to introduce an infrared regulator
(like a mass) with a parameter $m$, respecting the symmetry of the
system; at the end of the computation one takes the limit $m
\rightarrow 0$.  The reason for  introducing a regulator is that,
due to the parity anomaly one cannot consistently define a
gauge-invariant coupling for massless Dirac fermions in 2D. The mass
regulator breaks parity and even after it is sent to zero, in the
gauge-effective action its remnant is still there, keeping the
information of the mass sign in the coefficient of the generated
Chern-Simons action.  However,  for our system one cannot take as
regulator simply a mass term in the lattice as in the standard
systems, since it would preserve the time-reversal symmetry, broken
in our case.

% Let us denote the holon annihilation operators in the two sublattices,
% A(even sites) and B(odd sites) by $\hat h^a$ and $\hat h^b$,
% respectively.
A regularized free Hamiltonian for the field
$( h^a(\vec k), h^b(\vec k))^t$ maintaining the above discussed
symmetry has a matrix form in the momentum space given by:
\begin{eqnarray}
\label{eq:20}
&H(\vec k)= \nonumber \\ &\begin{pmatrix}
 m \cos (k_x+k_y)&2t (\cos k_x + i \cos k_y)\\ 2t (\cos k_x-i \cos k_y)&m \cos (k_x-k_y)
\end{pmatrix}.
\end{eqnarray}
One can check directly that this Hamiltonian with the regulator
added respects the combined symmetry of time reversal and the
exchange of N\'{e}el lattice of the original Hamiltonian. To be
specific the time reversal operation is implemented by complex
conjugation and $\vec k \rightarrow -\vec k$, while the interchange
of
 N\'{e}el sublattices corresponds to $k_y \rightarrow - k_y$ ( a
mirror reflection about the x-axis ) followed by a similarity
transformation implemented by $\sigma^x$. In our units the Hall
conductivity of the lower bands is given by $c_1/(2\pi)$, where
$c_1$ is the Chern number of the corresponding bands. For a
two-dimensional $H(\vec k)$ as ours, $c_1$ can be computed as
follows (See, $e.g.$, Ref. \onlinecite{st}): We write $H(\vec k)$ in
terms of $ \sigma^\mu=({ \bf 1},\vec \sigma)$, with $\mu=0,1,2,3$
and $\vec \sigma$ the Pauli matrices: $H(\vec k)= \sum_\mu
H_\mu(\vec k)\sigma^\mu$. We call $D$ the set of points in the BZ
where $H_1=H_2=0$, which are called Dirac points, then
\begin{eqnarray}
\label{eq:20.1}
c_1=\frac{1}{2} \sum_{x \in D} {\text{sign}}[ H_3(x)]{\text{sign}}[
 \epsilon_{3ij} \frac{\partial H_i}{\partial k_1}  \frac{\partial H_j}{\partial k_2}(x)].
\end{eqnarray}

If we compute the Chern number $c_1$ of the lower bands of $H(\vec
k)$, describing ``spurious'' holons as discussed above, one then
finds 1, since at the two Dirac points (becoming for $m=0$ the Dirac
nodes) the regulator term has opposite sign :
$H_3(-\pi/2,\pi/2)=m,H_3(\pi/2,\pi/2)=-m$ and the second sign in
Eq.~(\ref{eq:20.1}) is also opposite at those points.  We then see
that the lower bands in our holon system contribute $
\text{sgn}(m)/(2 \pi)$ to the Hall conductivity.

In order to discuss the spinon coupling to
%of
the ``physical'' upper bands taking into account the gap of spinons,
as done in the next subsection, one needs to go to the
long-wavelength continuum limit. In that limit the lower bands of
the above Hamiltonian are just the lower bands of two Dirac
double-cones regularized with the same mass $m$. Since every cone
contributes to the Hall conductivity with $ m /(4 \pi
|m|)$,\cite{re} we see again that the lower bands in our system
contribute $ \text{sgn}(m)/(2 \pi)$.

\subsection{Hall conductivity of ``physical'' holons}

In the absence of the spinon coupling, for free Dirac holons the
partially filled upper bands would contribute exactly the opposite
Hall conductivity of the lower bands, since in case of partial
filling the Hall conductivity of the free system is zero.
Introducing a mass, we get a non-vanishing result only if the
chemical potential is in the mass gap, which is not our case.
However, the coupling of the upper band to spinons changes the
situation.

To discuss the effect of spinon coupling we need to extract an
effective action and compute the coefficient of the corresponding
Chern-Simons term for the $b$ field. This calculation involves a
mixing of the upper ``physical'' and lower ``spurious''
  bands. In order to minimize such  mixing, a careful
  treatment is needed. In fact, we need only consider the mixing in an
  infinitesimal neighborhood of the Dirac nodes following the procedure
  outlined below; more details are deferred to the Appendix.

We consider one partially filled Dirac double-cone, while the other
one can be treated in the same way. To identify the Green function
of the continuum fields associated with the two bands we start with
rewriting the relevant free Dirac propagator $G$ with chemical
potential $\mu_F$ at $T=0$ in the following form: let $\gamma^\mu$
with $\mu =0,1,2$ denote the 2+1 Dirac gamma matrices and $k^\mu =
\omega, \vec k$ the 3-momentum.

Then we find\cite{su},:
\begin{eqnarray}
\label{eq:20b}
&G= (\slashed{k}-m) [\Theta(-k^0) \frac{1}{k^2-m^2 + i \epsilon}+\nonumber \\& \Theta(k^0)(\frac{\Theta(k^0-\mu_F)}{k^2-m^2 + i \epsilon}+\frac{\Theta(\mu_F-k^0)}{k^2-m^2 - i \epsilon})].
\end{eqnarray}
Naively the first term corresponds to the lower band, but to take
into account the problem of mixing quoted above we extend the first
term up to a small cutoff $\eta$ with $\mu_F \gg \eta \gg |m|$ to
include the bottom of the upper band, replacing $\Theta(-k^0)$ by
$\Theta(-k^0+ \eta)$ and subtracting the corresponding contribution
in the second term of Eq.~(\ref{eq:20b}). Note that $\eta$ is
  eventually sent to zero, after the limit $m \rightarrow 0$ has been
  taken.  Hence the introduction of the small cutoff $\eta$ takes into
account only the contribution from the conduction band edge. Since
the relevant contribution for physical holons at large scales comes
only from the region near the Fermi surface, it is unmodified by the
above operation.  According to the previous discussion we then
insert in the modified second term (assumed to describe the
``physical'' holons) the minimal coupling to $b$, to  spinons $s$
and to $v$, whereas we insert only the minimal coupling to $b$ in
the modified first term describing the ``spurious'' holons appeared
with the violation of the Gutzwiller projection.

We have already calculated above the Hall conductivity of the first
term, $\sigma_H = 1/(2 \pi)$; correspondingly the leading
contribution in the long wavelength continuum limit of the effective
action is given by the Chern-Simons term $S_{c.s.}^{U(1)}(b)$. We
now discuss the Hall conductivity of the second term.

The long wavelength continuum limit of the spinon interaction is
just the minimal coupling of  holons to the slave-particle gauge
field $A_\mu(x) \sim s^*(x) \partial_\mu s(x)$. This is the gauge field of the CP${}^1$
representation of the O(3) spinon $\sigma$ model, implementing in the
continuum the slave-particle gauge invariance.
%(see Eq. (
%\ref{eq:13})).
Then the leading term of the effective action due to
``physical'' holons turns out to be  $-S_{c.s.}^{U(1)}(b +A)
-S_{c.s.}^{U(1)}(v)$, as shown in the appendix.

We now need to integrate $A$ to find both Hall conductivity and, as
required by Eq. (\ref{eq:18}), the compressibility of the holon
system. The compressibility is proportional to the scalar
polarization bubble evaluated at zero energy in the limit of zero
momenta.

Since the upper band is partially filled, the leading contribution
comes from a region near the Fermi surface. Then at $\omega =0$ in
the limit $k \rightarrow 0$ in the Coulomb gauge its polarization
bubble matrix is given by:
\begin{eqnarray}
\label{eq:21}
\pi^h(\vec k)=\begin{pmatrix}
  \chi_0^h  &  -k_2 \sigma_H^h &  k_1 \sigma_H^h\\  k_2 \sigma_H^h &  \vec k^2 \chi_\perp^h & 0 \\ -k_1 \sigma_H^h & 0 &  \vec k^2 \chi_\perp^h
\end{pmatrix},
\end{eqnarray}
where $\chi_0^h,\chi_\perp^h,\sigma_H^h=-1/(2 \pi)$ are the density of
states at the Fermi energy, the diamagnetic susceptibility and the Hall
conductivity, respectively.

As spinons are gapped, integrating them out one obtains a Maxwell
effective action for the slave-particle gauge field $A_\mu$; the
spinon polarization bubble matrix at $\omega =0$ is then given by
the diagonal matrix
\begin{eqnarray}
\label{eq:22}
\pi^s(\vec k)={\rm diag}( \chi_0^s k^2,  \vec k^2 \chi_\perp^s, \vec k^2 \chi_\perp^s),
\end{eqnarray}
with $\chi_0^s,\chi_\perp^s$ the electric and the diamagnetic
susceptibility of the spinon system.
The scaling analysis presented in
  Ref. \onlinecite{frogoma} based upon a tomographic representation of
  fermion Green functions suggests that the RPA approximation gives
  the leading term in the scaling limit of the polarization bubbles of
  holons in the presence of the Maxwell interaction originated from
  spinons, described by $(\pi^s)^{-1}$, since such interaction is of long-range. 
  The polarization bubble
of \emph{holons} dressed by the spinon interaction in RPA in the small $k$
limit is then given by
\begin{eqnarray}
\label{eq:25}
\Pi^{h}(\vec k)=  \pi^h[1+(\pi^s)^{-1}\pi^h]^{-1},
\end{eqnarray}
where the scalar component (corresponding to compressibility) reads
\begin{eqnarray}
\label{eq:23} \Pi^h_0(\vec k)=  \chi_0^s k^2,
\end{eqnarray}
while the Hall polarization bubble by
\begin{eqnarray}
\label{eq:24} \sigma_H(\vec k)=\frac{\sigma_H^h k^2 \chi_0^s
\chi_\perp^s}{(\sigma_H^h)^2+ \chi_0^h (\chi_\perp^s +
\chi_\perp^h)};
\end{eqnarray}
therefore both vanish at $k=0$, implying that the upper bands of
``physical'' holons are incompressible and do not contribute to the Hall
conductivity for $b$. The origin of incompressibility can be traced back to the
unscreened long-range 2D Coulomb repulsion generated by the slave-particle gauge field,
 due to the spinon gap. Hence it is a consequence of the Gutzwiller constraint, origin of the
  gauge field, and of the destruction of the N\'{e}el order due to the AF vortices introduced by  doping.

Since the lower holon bands are completely filled it then turns out
that the total holon system before it is coupled to the Chern-Simons
$b$-field is incompressible and provides Hall conductivity $1/(2
\pi)$.  Hence, according to the result stated at the beginning of
the Section, after coupling with $b$ the resulting semionic holons
have exclusion statistics parameter 1/2, as we would like to prove.

As seen in the proof, the role of the slave-particle gauge field, as
a direct consequence of the no-double occupation constraint, is
crucial to obtain the 1/2 exclusion statistics, exactly as in 1D
case, where the constraint is also crucial to realize the 1/2
  exclusion statistics by producing a Luttinger-type interaction.
  Notice that incompressibility of the holon liquid does not imply incompressibility of the hole liquid, because the polarization bubble of the hole  involves also the renormalization  of the gauge propagators due to the holons.

Furthermore, since the Chern-Simons term of $v$ does not contain a
coupling to $A$ one finds that the total Chern-Simons contribution
to the effective action of (both ``physical'' and ``spurious'')
holons in the bulk is given by $S_{c.s.}^{U(1)}(b)-S_{c.s.}^{U(1)}(v)$.
(Also a gauged Wess-Zumino-Novikov-Witten boundary term is generated by gauge
invariance \cite{wen},\cite{frs}.) Therefore, although broken by the system of semionic holons alone,
parity and time-reversal symmetries are still explicitly preserved in the
holon system couped to the spinon-inherited $v$ field. In fact, the bands of ``spurious'' holons
produce a chiral structure, due to $b$, but the bands of ``physical'' holons produce an opposite chiral
 structure, due to $v$.  The induced additional Chern-Simons terms change, however,  the braid statistics
 of the low-energy holon quasi-particle excitations (in Landau's sense) near the ground state of the
 semionic holon liquid. Since the coefficients of the total, original plus induced,  Chern-Simon actions
 for $b$ and $v$ are -1 and +1, respectively, we expect  that the statistics of such quasi-particles is
 fermionic, as indeed suggested by preliminary calculations based on an approximate explicit expression  for the low-energy behaviour of
 the holon Green function that will be presented in a separate paper.
% but to actually prove this requires an explicit expression for the low-energy behaviour of
 %the holon Green function, that is presently under investigation.
 This change of statistics from the
 fields to the "Landau" quasi-particle excitations is somewhat analogous to what happens in the composite
  fermion theory of the Fractional quantum Hall effect \cite{jain}, where the quasi-particle excitations
   near the composite fermion ground state are anyons.

%With a proper regularization, the role of the ``spurious'' holon
%bands is to give an approximate but self-consistent description of
%the Gutzwiller projection on holons, providing an effective
%``vacuum'' in which the ``physical'' hopping holons move. But this
%``vacuum'' is topologically non-trivial, as shown by its
%non-vanishing Chern number or Hall conductivity, and this
%non-triviality deeply affects the physics of the ``physical'' upper
%bands, in particular forcing a 1/2 exclusion statistics for its
%semions.

Above we discussed the situation in the PG ``phase'', now we add a
brief comment for the SM ``phase''. As quoted in Sect. 2, the
optimal spinon configuration $\tilde  s^m$, around which we expand the
spinon fluctuations described by $s$, acquires for hopping holons
$-\pi$-flux phase factors in the SM ``phase'' that cancel the original
$\pi$-flux phase factors. The spinon coupling
occurs only for the upper band and this additional phase factor
modifies the dispersion of the upper band so that close to the Fermi
  surface it is turned into that of the $t$-$J$ model in the
tight-binding approximation; both compressibility and the Hall
conductivity of the upper band still vanish. Since the lower band is
not involved in the modification its Hall conductivity remains $1/(2
\pi)$ , hence even in SM phase the hopping holons have exclusion
statistics 1/2.

On the basis of the result on the Fermi surface discussed in Sect. III
 and the generalization of the Luttinger theorem discussed in Ref. \onlinecite{sent},
 we suspect that the physical hole system, at least in the PG ``phase'' ,
 possesses a ${\bf Z_2}$ topological order of the kind considered in the above quoted references\cite{sent}.

\section{Conclusions}
Let's summarize our results. The Gutzwiller projection and the low
dimensionality (1 or 2D), allow a gauging of the $U(1)$-charge and
$SU(2)$-spin symmetries of the $t$-$J$ model leaving its physics
completely unmodified. As a result at the lattice level the charge
degrees of freedom, described by spinless holons, and the spin
degrees of freedom, described by spinons, of the $t$-$J$ model
acquire a semionic braid statistics both in 2D, and in 1D, where
this statistics holds also for the corresponding low-energy
"quasi-particles" and can be explicitly checked comparing with the
exact solution. The additional freedom provided by the $SU(2)$
gauging allows a better simultaneous optimization of both  $t$ and
$J$ terms, and in 2D it introduces a novel kind of excitations,
$i.e.$, the AF spin vortices. These are quantum distortions of the
AF spin background in the $U(1)$ subgroup of the $SU(2)$-spin group
unbroken by antiferromagnetism. Their cores are located on the empty
sites of the 2D $t$-$J$ model, mimicking the Zhang-Rice singlets of
cuprates, and they record in their vorticity the N\'{e}el structure
of the lattice. In a mean-field treatment at large scales their
interaction with the spin degrees of freedom turn the long-range AF
order of the model at half-filling into a short-range AF order above
a critical doping and this implies a relaxation of the Gutzwiller
constraint at large scales. Although with the Gutzwiller projection
exactly implemented the charge degrees of freedom have physical
bands empty at half-filling, at mean field level the constraint is
relaxed and ``spurious'' filled lower holon bands appear, describing
the unprojected holes at half-filling.  With a proper
regularization, the role of these ``spurious'' holon bands is to
give an approximate but self-consistent description of the
Gutzwiller projection on holons, providing an effective ``vacuum''
in which the ``physical'' hopping holons move. But this ``vacuum''
is topologically non-trivial, as shown by its non-vanishing Chern
number or Hall conductivity, and this non-triviality deeply affects
the ``physical'' upper bands, in particular forcing a 1/2 exclusion
statistics for their semionic holons, as in 1D. The slave-particle
gauge attraction between holon and spinon then produces a Fermi
surface for  holes, as holon-spinon low-energy bound states, with
the addition of a $t'$-term, approximately consistent with the ARPES
experiments in hole-doped
cuprates \cite{mysy}.%, as proved previously, but only qualitatively, in an approximate treatment where the 1/2 exclusion statistics was assumed.
As in 1D the parity and time-reversal symmetry are broken separately
for the holon and spinon subsystems, due to their semionic nature, but
for physical slave-particle gauge-invariant quantities they are restored.

To conclude, we remark that the spin-charge gauge approach allows to
recover, sometimes even semi-quantitatively, many unusual
experimental features of hole-doped cuprates, and for completeness
we now briefly mention the most relevant results. We believe that
the most interesting feature of the approach is that  holes are
composites made of only weakly bound holons and spinons,  so that
some physical responses are dominated by the spin carriers, in
totally non-Fermi liquid manner.  As noticed above, previously this
approach was implemented in the approximation in which the 1/2
Haldane statistics for the holon was assumed, but, somewhat
inconsistently, its semionic nature was not taken into account,
consistently neglecting, however, the influence of AF spin vortices on holon
hopping. The expected fermionic nature of the "Landau" holon
quasi-particle discussed in the previous section nevertheless
suggests that the approximation made was already not unreasonable.
Anyway a careful account of the two neglected effects mentioned
above cannot change significantly the physical responses dominated
by  spinons, which are responsible for  results that we now sketch
in words, referring to the original papers for explicit formulas and
plots.

A phenomenon naturally explained by this approach is the
metal-insulator crossover (MIC) found decreasing $T$ in the in-plane
resistivity of underdoped cuprates (see $e.g.$ Refs.
\onlinecite{MIC}). Although it is often attributed to
disorder-induced localization, that interpretation is at odds with
the fact that, depending on materials and dopings, MIC occurs from
far below to far above the Ioffe-Regel limit. That interpretation is
also at odds  with the existence in a large range of temperatures,
including the MIC, of a universal curve\cite{unir} for a normalized
resistivity as a function of $T/T^*$, where $T^*$ can be identified
as an inflection point in the in-plane resistivity. For these
reasons we believe that the MIC is intrinsic, although
disorder-induced localization may play a role at lower temperatures
where, in fact, universality breaks down. In the spin-charge gauge
approach the MIC can be easily explained: Due to the slave-particle
gauge string binding spinon to holon, the velocity of the hole bound
state is determined by the slowest among spinon and holon
(Ioffe-Larkin rule \cite{il}). The holon has a metallic behavior
with a FS, whereas,
%so the hole exhibits a
%metallic/insulating crossover if the spinon does.
due to the AF gap, at low $T$ the spinon can only move by thermal
diffusion leading to a semiconducting behavior. However, at higher
temperatures its dynamics is dominated by the dissipation growing
with $T$ induced by slave-particle gauge fluctuations, leading to a
metallic behaviour. The universality is explained by the spinon
dominance \cite{msy}, leading to insensitivity to details of the FS,
and even quantitatively the universal curve can be well reproduced
\cite{mb}. We call $T^*$ the low-pseudogap temperature at which the
resistivity curve exhibits an inflection point and, on the basis of
the comparison between the experimental and the theoretically
derived resistivity curve, we identify it with the crossover from PG
to SM in our approach. A similar crossover with increasing $T$ from
a AF-gap dominated region to a gauge-induced dissipation dominated
region can explain the peak in the spin-lattice relaxation rate
${}^{63}(1/T_1T)$ in underdoped cuprates \cite{slr}.

The spin-charge gauge approach provides a three-step mechanism for superconductivity
 that might explain several crossovers appearing in the phase-diagram of cuprates.

 Firstly, at a temperature that we denote by $T_{ph}$, the attraction mediated by the
  AF-spin vortices described in  Eq.~(\ref{eq:15}) produces  charge-pairing, the spin
  degrees of freedom being still unpaired. Since
the formation of charge-pairs induces a reduction of the spectral
weight on the FS of  holons, inherited then by  holes, we identify
in cuprates $T_{ph}$ as the temperature below which a pseudogap
appears in the spectral weight of the hole (even well above $T^*$ in
the $t$-$t'$-$J$ model \cite{mg}), and we call this temperature
high-pseudogap. Qualitatively many features of this high-pseudogap
in the hole spectral weight derived in Ref. \onlinecite{mg} are
consistent with experimental data, but, according to the result of
the present paper the influence of the semionic nature of the holon
field should be reconsidered.

At a lower temperature, $T_{ps}$, the slave-particle gauge
attraction between holon and spinon induces the formation of
short-range spin-singlet (RVB) spinon pairs, in a sense, using the
holon-pairs as the source of attraction, thus leading to a finite
density of incoherent hole pairs. Comparing the behavior of
spinon-pair density \cite{mfsy},\cite{m} with the intensity of the Nernst signal \cite{ner} seen in
cuprates, we identify this crossover as the onset of the diamagnetic/Nernst signals
induced by magnetic vortices.

Finally, at an even lower temperature, the superconducting
transition temperature $T_c$, the hole pairs become coherent and a
$d$-wave hole condensate appears, leading to superconductivity. The
presence of three crossover temperatures $T^*,T_{ph},T_{ps}$ is
typical of this approach and finds a reasonable correspondence in
the experimental phase diagram of cuprates\cite{m}.

In particular, for the same reason given for in-plane resistivity,
the superfluid density satisfies Ioffe-Larkin rule and is dominated
by  spinons in the underdoped region. Below $T_{ps}$ the low-energy
effective action obtained integrating out the massive spinons is a
Maxwell-gauged 3D XY model, where the angle-field of the XY model is
the phase of the long-wave
 limit of the hole-pair field and the gauge
field is the slave-particle gauge field.% Below $T_c$ the gauge field is gapped by the Anderson-Higgs mechanism, the superconducting
%transition therefore is almost of classical 3D XY type,
 This explains the 3DXY critical exponent of the superfluid density found in  experiments.
 Furthermore, as in the case of resistivity, the spinon dominance explains the experimental
 observation of a universal curve for the normalized superfluid density as a function of $T/T_c$ \cite{unis},
  which  can be well reproduced even quantitatively by the spin-charge gauge approach\cite{mb}.

Let us end this paper by remarking that we 
are computing
%plan to compute
 some
physical response dominated by holons, to check the effect of the
semionic nature of the holon field in our approach, in comparison
with the experimental data of cuprates.
Preliminary calculations suggest that the main effect w.r.t. the previous approximate treatment is a modification of the wave-function renormalization constant of the hole, which becomes temperature independent allowing, for example, a recovery of  the experimentally observed Fermi-liquid behaviour of the Knight shift at high $T$ in the "strange-metal phase" of hole-doped cuprates (see e.g. Ref. \onlinecite{sli}).

%The combined semionic braid statistics of holons and
%spinons together with the 1/2 exclusion statistics of holons and the
%charge-pairing induced by antiferromagnetic spin vortices allow then
%to recover, sometimes even semi-quantitatively, many experimental
%features of the phase diagram of the cuprates.

\section*{Acknowledgements}
We thank the Referees for stimulating comments that partially
motivated the preparation of Ref. \onlinecite{mysy}. F.Y. is
supported by National Nature Science Foundation of China 11774143,
and JCYJ20160531190535310.  P.A.M. acknowledges the partial support
from the Ministero Istruzione Universit\'a Ricerca (PRIN Project
"Collective Quantum Phenomena: From Strongly-Correlated Systems to
Quantum Simulators").

\section{Appendix}

In this Appendix we outline the calculation of the Chern-Simons term
for the upper band of holons in one of the Dirac double-cones. For
simplicity we consider only the coupling to $a \equiv A+b$, the
coupling with $v$ can be done in a similar way. The two terms don't
mix due to the $\sigma^z$ factor in the $v$-coupling arising from
the fact that the two components of the Dirac field arise from
different N\'{e}el sublattices, hence with opposite charge for $v$.
Following Ref. \onlinecite{re} we consider a coupling to a field $a$
of constant field strength $f_{\mu \nu}$, calculate the expectation
of the induced current
\begin{eqnarray}
\label{eq:A1}
\langle J^\mu \rangle =  \mathrm{Tr}  [\gamma^\mu G_a(x,x')]_{x \rightarrow x'}
\end{eqnarray}
with $G_a$ the gauge invariantly regularized Green function in the
presence of $a$, and we keep only the term proportional to
$\epsilon^{\mu \nu \rho} f_{\nu \rho}$. Its coefficient, which we denote by
$J_{CS}$, multiplied by $4 \pi$ is the coefficient
of the Chern-Simons action. We define $K_\mu = k_\mu + a_\mu$ and
consider the case $a_0=0$.  According to the discussion in sect. III D,
the Green functions for free "physical" holons in momentum space is
given by
\begin{eqnarray}
\label{eq:A1.1}
&G= (\slashed{k}-m) \Theta(k^0)[(\frac{\Theta(k^0-\mu_F)}{k^2-m^2 + i \epsilon} \nonumber +\frac{\Theta(\mu_F-k^0)}{k^2-m^2 - i \epsilon})\nonumber \\ &- \Theta(\eta-k^0) \frac{1}{k^2-m^2 + i \epsilon}],
\end{eqnarray}
with $\mu_F \gg \eta \gg |m|$. Then, using Schwinger's proper time
formalism, the Green function $G_a$ for the "physical" holons in the
limit ${x \rightarrow x'}$ can be represented as:
\begin{widetext}
\begin{eqnarray}
\label{eq:A2}
&[G_a(x,x')]_{x \rightarrow x'}= -i \int \frac{d^3k}{(2 \pi)^3} (\slashed{K}-m)
\int_0^\infty d s \Theta(k^0)[\Theta(k^0-\mu_F) e^{i s (k_0^2-m^2)} e^{-i s \vec K^2}e^{i s [\gamma^\nu,\gamma^\rho] f_{\nu \rho}/2}-\nonumber\\
&\Theta(\mu_F-k^0) e^{-i s (k_0^2-m^2)} e^{i s \vec K^2}e^{-i s [\gamma^\nu,\gamma^\rho] f_{\nu \rho}/2}-\Theta(\eta-k^0) e^{i s (k_0^2-m^2)} e^{-i s \vec K^2}e^{i s [\gamma^\nu,\gamma^\rho] f_{\nu \rho}/2}].
\end{eqnarray}
\end{widetext}
The relevant term for the Chern-Simons action in $\mathrm{Tr} ( \gamma^\mu (\slashed{K} -m) e^{i s [\gamma^\nu,\gamma^\rho] f_{\nu \rho}/2})$ is given by $ m s  \epsilon^{\mu \nu \rho} f_{\nu \rho}$. Inserting this term in (\ref{eq:A1}),(\ref{eq:A2}) and performing the integral over spatial momenta one finds
\begin{widetext}
\begin{eqnarray}
\label{eq:A3}
&J_{CS}= \frac{m}{8 \pi^2} \int d k^0 \int_0^\infty ds \Theta(k^0)(\Theta(k^0-\mu_F) e^{i s (k_0^2-m^2)}-\Theta(\mu_F-k^0) e^{-i s (k_0^2-m^2)}-\nonumber \\
&\Theta(\eta-k^0) e^{i s (k_0^2-m^2)})=\frac{1}{16 \pi}\frac{m}{|m|}(\Theta(|m|-\mu_F)-\Theta(\mu_F-|m|) -\Theta(\eta-|m|) + \frac{ 2 i}{\pi} \int_0^{\eta/|m|} P(\frac{1}{x^2-1})dx ),
\end{eqnarray}
\end{widetext}
where $P(\cdot)$ denotes the principal value. In the limit $m
\rightarrow 0$ the imaginary term disappears and one recovers the
result $-\frac{1}{8\pi}\frac{m}{|m|}$. A similar calculation can be
done for the band of "spurious" holons with free Green function
defined, as discussed in sect. III D, by
\begin{eqnarray}
\label{eq:A4}
G= (\slashed{k}-m) [\Theta(\eta-k^0) \frac{1}{k^2-m^2 + i \epsilon}]\nonumber.
\end{eqnarray}

Analogously one obtains, besides an imaginary term vanishing in the limit $m \rightarrow 0$,
\begin{eqnarray}
\label{eq:A5}
J_{CS}=\frac{1}{16 \pi}\frac{m}{|m|}(\Theta(\eta-|m|)+\Theta(\eta+|m|)) = \frac{1}{8\pi}\frac{m}{|m|}\nonumber,
\end{eqnarray}
proving that indeed the lower band of "spurious" holons has the Hall conductance calculated in
sec. IIIC.

\end{document}